# Small footprint optoelectrodes for simultaneous readout and passive light localization by the use of ring resonators

*Vittorino Lanzio[1,5]\*, Gregory Telian[3], Alexander Koshelev[2], Paolo Micheletti[5], Gianni Presti[1], Elisa D'Arpa[5], Paolo De Martino[5], Monica Lorenzon[1], Peter Denes[4], Melanie West[1], Simone Sassolini[1], Scott Dhuey[1], Hillel Adesnik[3], Stefano Cabrini[1]\**

[1] The Molecular Foundry, Lawrence Berkeley National Laboratory, Berkeley, CA 94720, United States of America

[2] aBeam Technologies, Hayward, CA, 94541, United States of America

[3] Adesnik Lab, University of California Berkeley, Berkeley, CA 94720, United States of America

[4] Engineering Department, Lawrence Berkeley National Laboratory, Berkeley, CA 94720, United States of America

[5] Department of Applied Science and Technology, Politecnico di Torino, Torino, 10129, Italy

\*vittorinolanzio@lbl.gov, \*scabrini@lbl.gov

# Abstract

Neural probes are *in vivo* invasive devices that combine electrophysiology and optogenetics to gain insight into how the brain operates, down to the single neuron and its network activity. Their integration of stimulation sites and sensors allows for recording and manipulating neurons' activity with a high spatiotemporal resolution. State-of-the-art probes are limited by tradeoffs between their lateral dimension, the number of sensors, and the ability to selectively access independent stimulation sites.

Here, we realize a highly scalable probe that features a three-dimensional integration of small-footprint arrays of sensors and nanophotonic circuits and scales the density of sensors per cross-section by one order of magnitude with respect to state-of-the-art devices. For the first time, we overcome the spatial limit of the nanophotonic circuit by coupling only one waveguide to numerous optical ring resonators as passive nanophotonic switches. With our strategy, we achieve accurate on-demand light localization while avoiding spatial-demanding bundles of waveguides and demonstrate the feasibility of a proof-of-concept device and its additional scalability, towards high-resolution and low-damaging neural optoelectrodes.

# Introduction

Exploring the human brain represents a multi-disciplinary challenge that has sparked a great deal of academic and industrial research [1] in recent years, aimed at understanding how information is processed and results in mental functions and behavior [2], as well as better understanding a number of diseases such as Parkinson's disease and other neurological disorders in the hope to develop new treatments [3].

Neurotechnology devices allow researchers to record and manipulate neural activity at different orders of magnitude in terms of interface volume (ranging from the whole brain to a few tens of neurons), spatial resolution (from a few cm to a few µm) and temporal resolution (from hours to sub-ms) [4], [5]. One effective way to interface with neurons with a high spatial and temporal resolution, on the order of single neurons and individual neural events, is to use invasive devices such as Michigan type neural probes, which utilize thin needles that integrate a high density of sensors for neural activity readout or electrical stimulation sites for neural activity manipulation in the needle's surrounding volume[6].

Michigan probes integrate arrays of metallic electrodes that record neurons' extracellular potentials and enable the triangulation of the neurons' positions by knowing the timing and amplitude of the signals on each electrode [7]. Such electrode arrays can also be used to stimulate neurons at the cost of both interfering with the electrophysiological recordings and the impossibility of targeting specific types of neurons [8]. With the development of optogenetics [9] it is now possible to simultaneously record neurons electrically and stimulate specific subpopulations optically [10], [11]. Previously, microLEDs [8],[9] or small waveguides have been used to optically stimulate cells that express optogenetic opsins [12], [13]. These light sources are typically placed above the surface of the brain for non-invasive manipulations or implanted to sub-cortical brain regions where non-invasive stimulation techniques do not work. This results in a fast and cell-type selective manipulation of neural circuits using light [14].

The combination of electrical readout and light stimulation is of great interest [15] to implement efficient feedback loops [16], [17]. Several state of the art neural probes allow for the combined recording of neural activity and simultaneous light stimulation on the same device but they all have slight drawbacks. For example, F. Wu et Al. integrate both electrodes and micro light-emitting diodes (µLEDs) for optogenetic stimulation [18], [19]. Here, µLEDs can be turned on and off independently, but cause heat generation during the µLED switching. Other approaches involve the use of waveguides for light delivery in addition to electrodes. Komal et Al. [20] fabricate probes with a Silicon Oxynitride waveguide and 8 electrodes per tip. Although this approach allows for multiwavelength illumination, it also presents scalability issues due to the bulky waveguide (30 um X 7 um) and does not allow for spatially conveying light in different device areas. Pisanello et Al. realize a tapered optical fiber able to selectively illuminate areas of interest but show no integration of electrodes [21], [22]. Two other groups tried to address some of these limitations by using integrated Si3N4 photonic circuits [12], [13]. Segev et Al. [12] fabricate a neural probe that allows one to select the light output location in a passive fashion, by using arrayed waveguide gratings. However, such a probe does not integrate any recording site. On the other hand, E. Shim et Al. [13] show a neural probe with nanophotonics circuits which allows for multi-point illumination coupled to recording electrodes, but does not have the ability to selectively control the spatial location of light. One more step towards this direction was reported by S. Libbrecht et Al. [23], who describe a neural probe integrating both electrodes and optical stimulation sites which can be activated independently. However, despite the important achievement of local stimulation of the neural network, this approach shows limitations in scalability because every light output requires a different waveguide. A recent optoelectrode enabling the selection of the light output location thanks to electrical switches demonstrated in vivo manipulation of single neurons [24]; however, the device is bulky and, similarly to other approaches, requires one waveguide on the device tip for multiple outputs. Based on these previous results, it emerges that, in order to investigate neural networks with high spatiotemporal resolution, an ideal device that combines both readout and manipulation of neural activity needs the following features: (i) the capability of addressing individual stimulation sites for accurate spatial control and (ii) the integration of a high number of sensors to simultaneously interface multiple neural cells [17]. In addition, such an ideal probe requires a minimization of tissue damage by both (iii) reducing implant size [25] and (iv) avoiding any heat generation [12], which can be achieved by employing a passive device, i. e., one that does not require any electrical current for its activation.

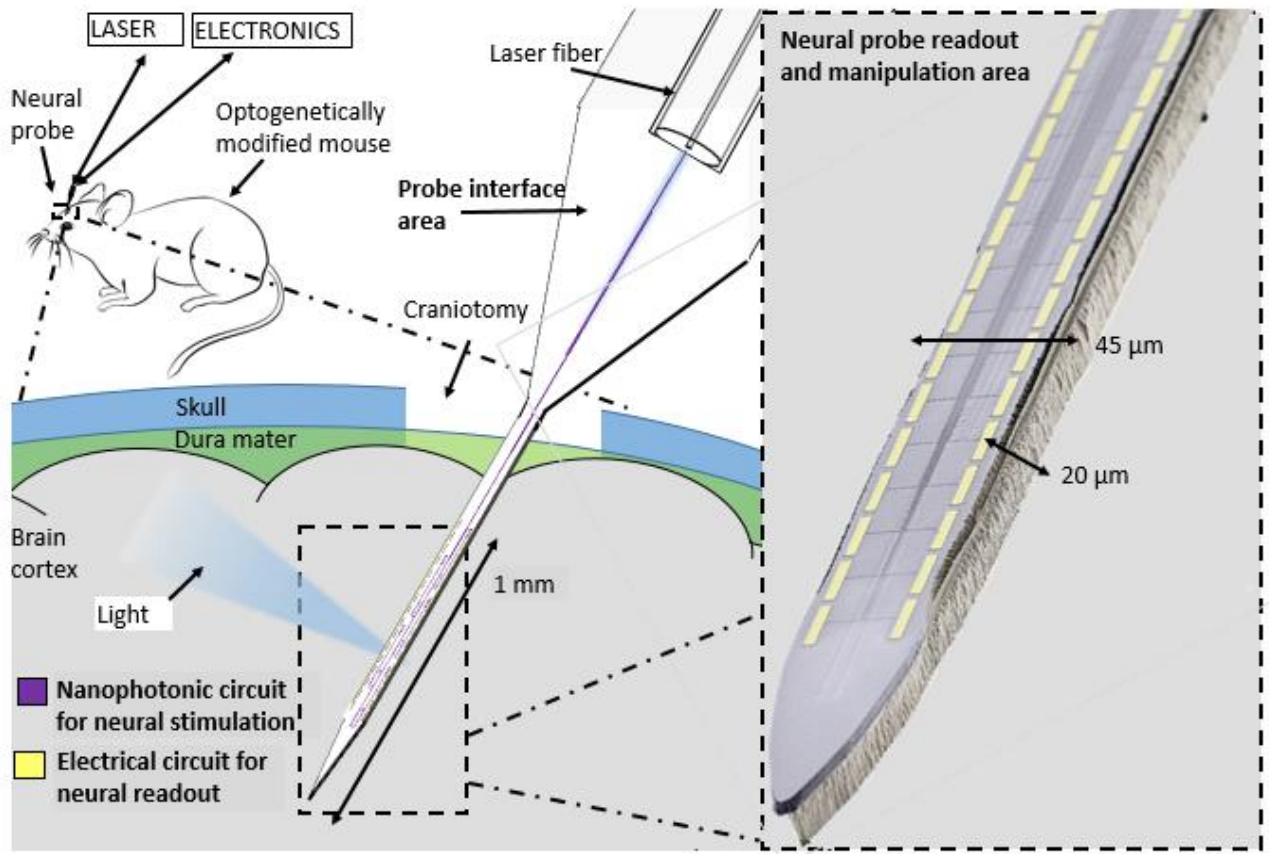

Figure 1. (a) Schematic illustration of the neural probe, which is a tiny needle device that is inserted into the brain cortex of an optogenetically modified mouse with the purpose of studying neural functions. The part of the device we insert into the cortex (the probe tip) has a minimally invasive size for reduced brain damage and integrates both nanophotonic circuits and arrays of sensors with a high density of sensors for simultaneous high spatiotemporal resolution optical manipulation and readout of neural networks. To access the tip's circuit functions, we connect the probe to a laser and external electronics by means of the device's interface area.

Several state of the art devices address one or more of these features but, to the best of our knowledge, no device combining both electrodes and passive nanophotonics enables for spatially controlling the light emission location while providing a reduced footprint and a high density of sensors. Here, we realize a neural probe (schematically illustrated in Figure 1) that allows for the simultaneous electrical readout and spatially addressable illumination. Our device integrates small footprint arrays of sensors and nanophotonic circuits on two different layers, thus preserving a high density of sensors while keeping small tip dimensions. With this strategy, we achieve a cross-sectional area coefficient (namely, the ratio between the tip cross-section and the total number of sensors and stimulation sites) of 12, which is one order of magnitude lower than state-of-the-art optoelectrical probes [23]. In addition, we report for the first time the use of ring resonators in the nanophotonic circuits for passive on-demand optical stimulation. To demonstrate the feasibility of a proof-of-concept device, we perform all the stages of design, fabrication, and characterization, followed by preliminary *in vivo* testing. We show that our strategy, which integrates both arrays of sensors and nanophotonic circuits embedding ring resonators, effectively combines all the ideal features for optoelectrodes: implant size reduction, increase of the number of sensors and stimulation sites, and light localization without heat generation.

## Results
### Optoelectrical neural probes' architecture
Our optoelectrical neural probes integrate both arrays of sensors for neural activity readout and nanophotonic circuits for passive and on-demand stimulation of the areas of interest. To realize such probes we combine micro and nanofabrication techniques, to optimize the device reproducibility and achieve high throughput and scalability. The probe (schematically illustrated in Figure 1) consists of three parts: *(i)* the tip, which is the electrical readout and stimulation area (the only part that we insert in the mouse's cortex), *(ii)* the interfacing area, which connects the tip's circuits to the external laser and electronics and *(iii)* a connecting area, which brings the electrical and optical signals from the tip to the interface area and vice-versa.

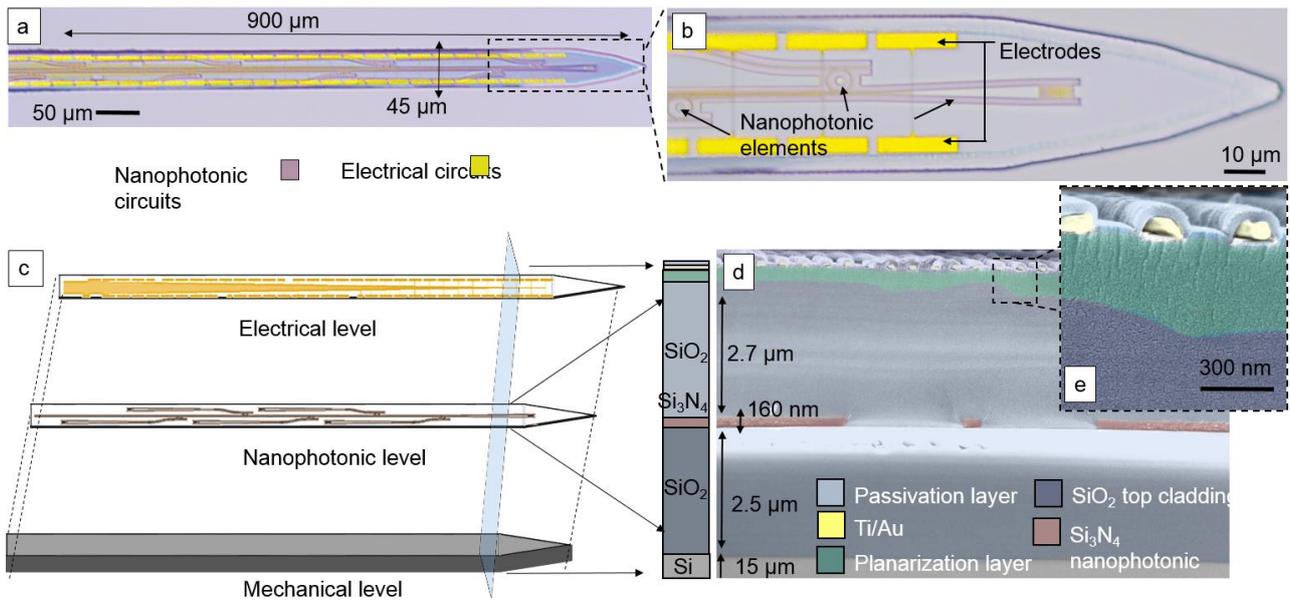

Figure 2. (a) Optical microscope top view image of the optoelectrical neural probe's tip, showing the integration of both the arrays of sensors (yellow areas) and nanophotonic circuits (light brown areas). (b) Close-up optical image of the top of the tip, displaying the closely spaced electrodes and the nanophotonic elements. (c) Schematic illustration of the three levels (mechanical, nanophotonic, and electrical) embedded in the tip. (d) False-color scanning electron microscope (SEM) images of a cross-section of the tip, showing the three levels and (e) a close-up of the $SiO_2$ passivated wires, which we pattern above the $SiO_2$ top cladding planarized layer.

Along its length, the tip is 900 µm long, 45 µm wide and 20 µm thick, as shown in Figure 2a. We design small footprint circuits (as shown in more detail in Figure 2b) and integrate them onto multiple levels, to keep the tip width as narrow as possible, which is essential to reduce brain damage. Specifically, we choose to embed three different levels, which we schematically illustrate in Figure 2c. The bottom level is made of a 15 µm thick silicon layer that gives mechanical stability to the tip. On the top of the silicon, we fabricate the level embedding the nanophotonic circuit, which is composed of patterned silicon nitride, sandwiched between two silicon dioxide capping layers ($SiO_2$, 2.5 µm bottom and 3 µm top cladding), shown in the SEM cross-section image in Figure 2d. The purpose of the nanophotonic circuits is to deliver on-demand light in specific locations along the tip length. Importantly, these nanophotonic circuits are passive and switching is enabled through circuits' wavelength sensitivity and external wavelength control, thus not requiring any electrical current flowing through the optical elements, thus avoiding any heat generation other than that caused from light itself, which is negligible compared to electrical heating [26], [27].

Above the nanophotonic circuit, we fabricate the level embedding the arrays of sensors, which are made of titanium and gold. This layer, which has been previously described in [28], is comprised of open electrodes for reading neural activity as well as $SiO_2$-passivated wires for transferring the signal from the electrodes to the probe interface area. The design of the nanophotonic and arrays of sensors is described in the next section. Both circuits are connected to an external laser source as well as electronics through the probe's interface area, which comprises of a groove for an optical fiber and large electrodes for wire bonding. We report the steps of fabrication and assembly of the device in a dedicated section (Device fabrication and assembly).

**Nanophotonics layer design and realization**

Nanophotonic circuits allow for selective illumination of specific locations along the tip of the probe and therefore stimulate neighboring neurons of interest. We achieve a high selectivity of localized light output by routing the light confined in the nanophotonic circuits by means of ring resonators [29], which act as passive optical switches. Different resonators can be selected simply by tuning the laser input wavelength via minor shifts in the input laser wavelength (<1nm), with no need for electrical current flowing through the optical elements. In addition, rings have the advantage of a small footprint, as only one main waveguide (also referred to as a bus waveguide) is needed to interface with all of the ring resonators, as opposed to other configurations that require one waveguide for each output spot.

Our nanophotonic circuit, which we schematically illustrate in Figure 3a, consists of a bus waveguide, several ring resonators (which we place along the length of the tip at an optimized distance from the bus), and, for each ring, an output waveguide terminated by a grating [28]. When the input laser light from the bus matches the ring's resonant frequency, which for a given material and thickness is mainly a function of its radius, it resonates due to constructive interference and transfers to the output waveguide, where it is extracted by the grating. Based on this model, we design the rings with different radii such that they resonate with different input wavelengths in our range of interest, as shown as an example in Figure 3a. With this strategy, we can passively select each ring and its relative light output location.

We set the initial wavelength range for the rings' resonance frequencies according to the maximum absorbance of the light-gated opsin channelrhodopsin (ChR2, centered at 450 nm [30] and then restrict it according to our laser's tunable range, which is 3.4 nm for the laser model we select (QFLD-450-10S from QPhotonics). We then optimize the ring parameters (gap, width, and radii, shown in Figure 3b) using finite difference time domain simulations (FDTD, Lumerical) [31].

We choose the ring free spectral range (FSR), which is the wavelength spacing between two resonances pertaining to the same ring [29], to be 3.21nm, a value that is close to the laser's tunability range as to maximize the number of rings within such range. We set the ring's Q factor (~1600) to be on the same order of the laser's (860±20), such that it absorbs all of the laser's energy and maximizes the percentage of light transferred from the bus to the output waveguide at the resonant frequency (called transmittance). By combining these two parameters, one can calculate the finesse, which is the ratio between the rings' FSR and their full width at half maximum (FWHM) and corresponds to the number of independently addressable rings. In our design we obtain a finesse close to 4, with an average ring transmittance of 69.5 ± 2.9 %, which we display in Figure 3c, along with the opsin absorption (light grey area) and the laser spectrum (grey curve). The cross talk (the percentage of light leaking from the selected ring to the adjacent ring) is below 6%, which is comparable with other nanophotonic technologies [24]; it can be decreased by designing more spaced ring resonances. Furthermore, it is worth noticing that our system's finesse is limited by the FWHM and the tunability range of our laser and could be incremented to hundreds of individual channels by choosing a different laser source. [32] Importantly, such an increase in the finesse and consequently in the number of rings would not correspond to an enlargement in the tip lateral dimension, due to the reduced nanophotonic elements' size, in addition to the fact that only one bus is required to interface all the rings. We choose the distance between gratings to be 150 μm; however, our strategy can accommodate for much denser light output configurations depending on the design of interest (see Methods Section).

Overall, our design yields a nanophotonic circuit with a lateral footprint as low as 15 μm, thus meeting the desired features for a small tip size.

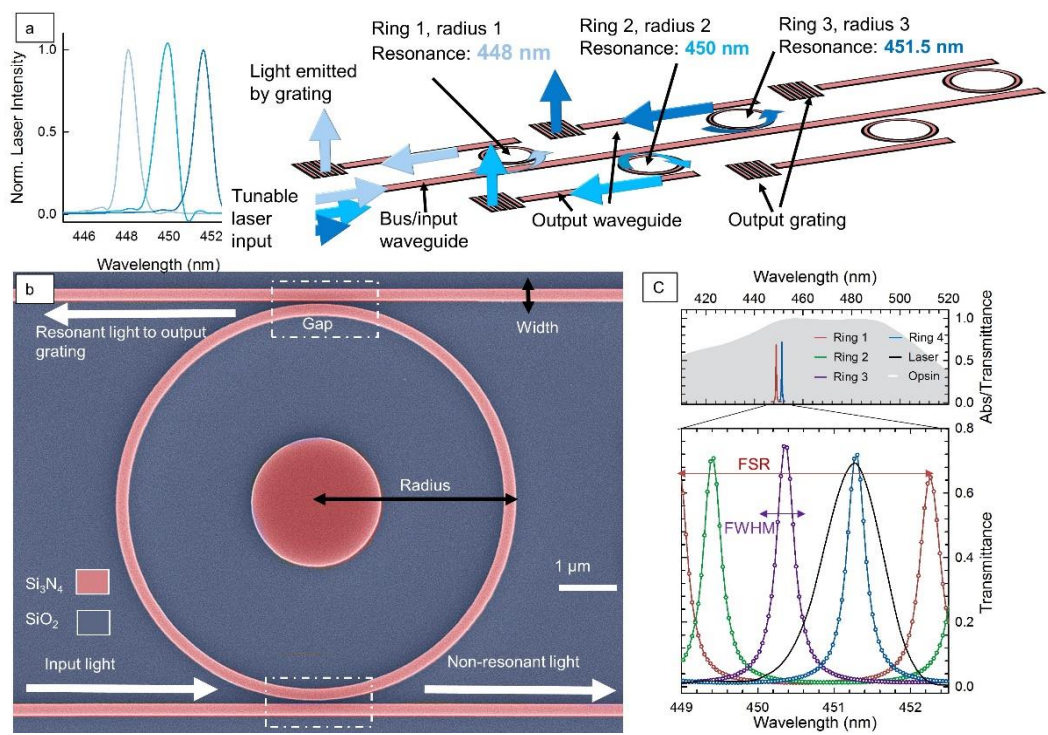

Figure 3. (a) Schematic illustration of the nanophotonic circuit in the probe tip area, composed of an input waveguide, ring resonators (with different radii and, therefore, different resonance wavelengths) and corresponding output waveguides and light extraction gratings. Arrows indicate the light path, with their color corresponding to the laser input spectrum on the left side. Nanophotonic circuits allow for the selection of the light output location: we show examples of this concept for the first three gratings, which we independently select by matching the laser input wavelength (with corresponding spectra on the left-hand side) to the desired ring's resonance frequency. (b) False-color SEM image of a ring resonator with a well-defined radius (3.881 μm), gap (80nm) and width (250 nm). We conceptualize the light propagation direction with white arrows, showing that the input light either

remains in the bus waveguide (if non-resonant with the ring) or transfers to the output waveguide (if resonant). (c) FDTD simulation of the ring transmittance (amount of light transferred from the bus to the output waveguide) for the implemented ring radii. All the ring resonances fit within the absorption peak of the ChR2 opsin (grey curve) and their FSR fits within the laser tunability range (graph boundaries). Ring resonance Q factors (resonance wavelength divided by FWHM [25]) are close to the laser one for coupling optimization (shown as a solid black line).

Above the nanophotonic layer we then integrate the arrays of sensors with the purpose of enabling a simultaneous readout of neural activity during light excitation. The readout circuit in the tip consists of 64 closely packed titanium/gold electrodes with a lateral dimension of 5 µm X 25 µm and pitch 27.5 µm (other electrode designs can be chosen according to the brain area of interest) and 64 corresponding electrodes in the probe's interface area. We connect pairs of electrodes at the two ends of the device with lithographically-defined metallic wires passivated by $SiO_2$, which are 120 nm wide and with 450 nm separation between them at the tip area and which widen up to 1 µm in the interface area.

The arrays of sensors must be defined on a planar surface to avoid wire collapse due to the severe roughness stemming from the presence of the $Si_3N_4$ nanophotonic elements. Hence we planarize the substrate with a 350 nm thick flowable oxide layer [33], as shown in Figures 2d-e. Importantly, this design enables one to place the arrays of sensors on top of the nanophotonics, which together with the limited wire width, prevents us from not exceeding the nanophotonic circuit's lateral footprint.

**Device fabrication and assembly**

We fabricate the nanophotonic and arrays of sensors and integrate them on the actual tips using micro and nanofabrication techniques, which allow us to obtain around 200 devices per wafer. The device fabrication process, sketched in Figure 4a, starts from a commercial silicon wafer with $SiO_2$ and $Si_3N_4$ optical quality layers [34] (Figure 4a). We initially pattern the nanophotonic circuits by electron beam lithography followed by dry etching and then clad them with 2.7 µm of PECVD deposited $SiO_2$ (Figure 4b), followed by the spinning and baking of flowable oxide (FOx 14 from Dow Corning) [33] in order to planarize the substrate. We then align and pattern the arrays of sensors with electron beam lithography, titanium/gold evaporation, and liftoff. After that, we proceed with the wire passivation by depositing 60 nm of $SiO_2$ with an atomic layer deposition tool; other thicknesses or materials could be used, but based on the material dissolution rate (< 1nm/day [35]) we expect to be able to use the probes for several chronic studies. We then remove the passivation layer from the electrodes by means of another electron beam lithography step and dry etching (Figure 4c), thus leaving the $SiO_2$ coating localized on the wires only. We chose electron beam lithography for convenience and design flexibility for the patterning of the nanophotonic circuits, the arrays of sensors and the passivation layer opening; however, other lithography techniques could be used for batch fabrication and higher throughput.

Next, we pattern the profile of the devices as well as the grooves for the alignment of the optical fiber from the wafer frontside (Figure 4d) by using optical lithography and dry etching. We finally release the devices from the wafer by a backside wet etching in potassium hydroxide solution (Figure 4e), which removes most of the silicon underneath the tip area (so to make it 20 µm thick) leaving behind the bulk silicon underneath the probe interfacing area. The device fabrication is fundamentally based on the processes described in better detail in [28].

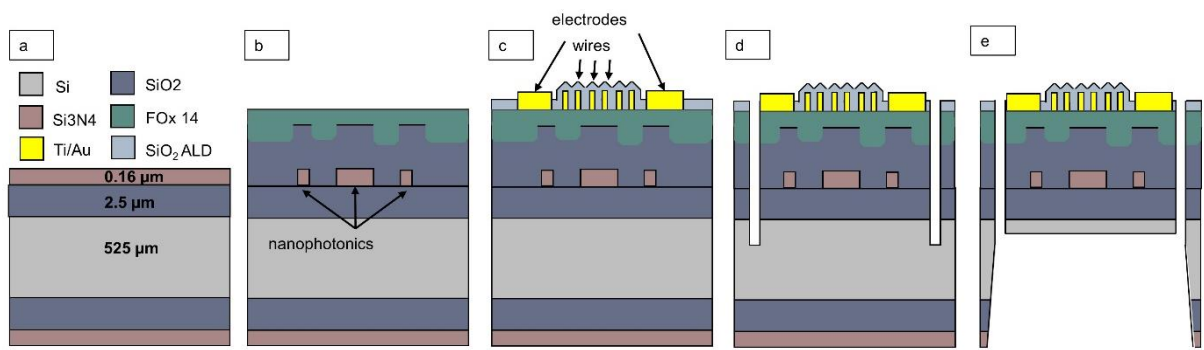

Figure 4. Sketch of the neural probe fabrication process. (a) Substrate wafer with $SiO_2$ and $Si_3N_4$ optical layers. (b) Patterning of nanophotonic circuits, optical insulation and planarization. (c)

Integration of the electronic circuits above the nanophotonic ones. (d) Patterning of device shape. (e) Release of the device by backside etching.

In order to be able to access the tip circuits and use their relative functions, we connect it to external electrical instrumentation and a laser source. Specifically, we achieve the electrical connection by gluing and wire bonding our probe on a custom made printed circuit board (PCB, shown in Figure 5a), which has on one side electrodes for wire bonding and on the other side electrical connectors. Moreover, we obtain the optical connection by coupling the laser's single-mode optical fiber to the edge of the bus waveguide (Figure 5b,c). We maximize the alignment between the fiber and the waveguide by using micro stepper motors while monitoring the probe's output, then dispense a low shrinkage, UV curable glue, and cure it through the optical fiber with 405 nm wavelength light to fix the fiber to the sample, so to secure the alignment.

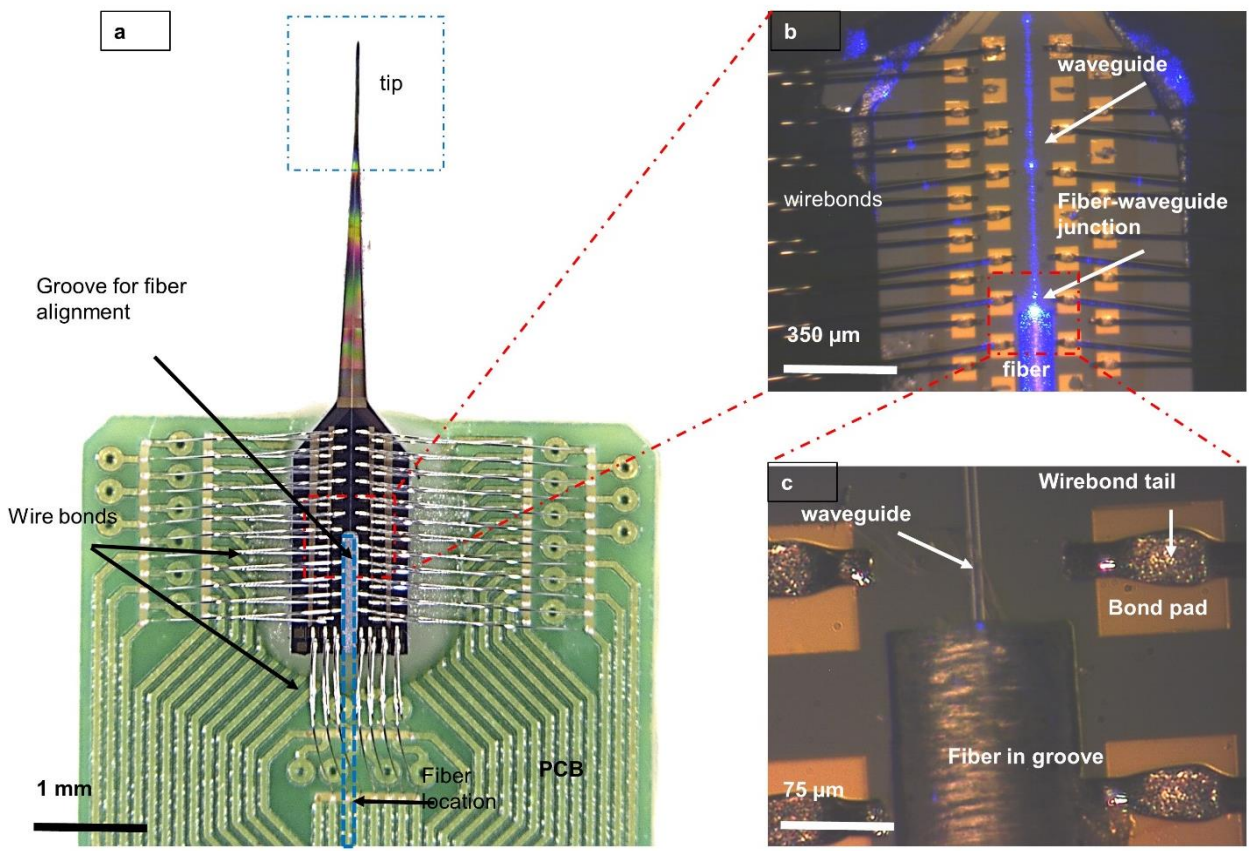

Figure 5: probe assembly steps to connect the nanophotonic and arrays of sensors to external instrumentation. (a): Device electrical connection. We show the whole device, comprised of the tip and of the interfacing area, which has bonding pads and an alignment groove for the optical fiber. The probe is glued on the PCB and wire bonded. (b): Device optical connection. We show a magnified image on the wire bonding and groove area after the optical fiber alignment. The laser is turned on and the waveguide (in blue) is visible due to sidewall scattering. (c): close up at the fiber-waveguide edge junction, showing in detail the fiber, the waveguide and the surrounding bonded pads.

**Optical characterization of the probe**

Once we have fabricated and assembled the neural probe, we proceeded to test its optical functionality in air to assess the nanophotonic circuit's capability of spatially addressing the light output location and to estimate the device's output power and losses. To do so, we connect the laser to the fiber that we have previously aligned and glued to the nanophotonic circuits' input and we monitor the tip's output gratings under an optical microscope, while simultaneously tuning the laser wavelength (within a range of 3.4nm)(more details in the Methods Sections). Moreover, we tune the laser's polarization with a paddle controller to match the polarization that we designed for the ring resonators (TE fundamental mode). We report an example of such a test in Figure 6, where we show a tip with five ring resonators. We demonstrate light spatial localization after (Figure 6a-d) we turn on the laser at the wavelengths corresponding to the first four ring resonances. A video of the light output switching is available in the Supplementary Information. We measure the rings' output transmittances and plot them in Figure 6e, along with the laser spectrum. From this measurement, we extract the experimental ring Q factor (on average: $861 \pm 127$), which closely matches that of the laser's ($860 \pm 20$). The rings' coupling efficiency, calculated as each grating's output intensity divided by the intensity of all the gratings, is between 45% to 60%, and the experimental cross-talk is on average $5.2 \pm 2.59\%$, with few rings measured in some of the probes having higher cross-talk probably due to some fabrication defects. The ring's output intensities show good uniformity with variation (calculated as standard deviation) of 11.3%. The ring's free spectral range is not available since it is wider than our laser's tunability range.

Finally, we evaluate the tip output power which, for a given laser input power, is limited by the device losses, which are mainly ascribed to fiber-waveguide coupling and waveguide scattering (see details in the Methods section). System loss minimization is crucial to output enough light power to be able to activate the opsin, which has been estimated to be around 1 mW/mm² [30].

Our laser diode has a maximum output power of around 10 mW, while the total system losses amount to ~30 dB, which leaves the total output power to be around 5 to 10 µW. This value corresponds to a power density up to a maximum of around 100 mW/mm$^2$ (considering a light output spot on the order of the grating dimensions, 5 um X 10 um), thereby showing that our system can output more than enough light for opsin activation (less light can be delivered by lowering the laser's input power), as we validate during the *in vivo* experiment (see next paragraphs). It is worth noting that the system losses can be drastically decreased by further optimization (see Methods).

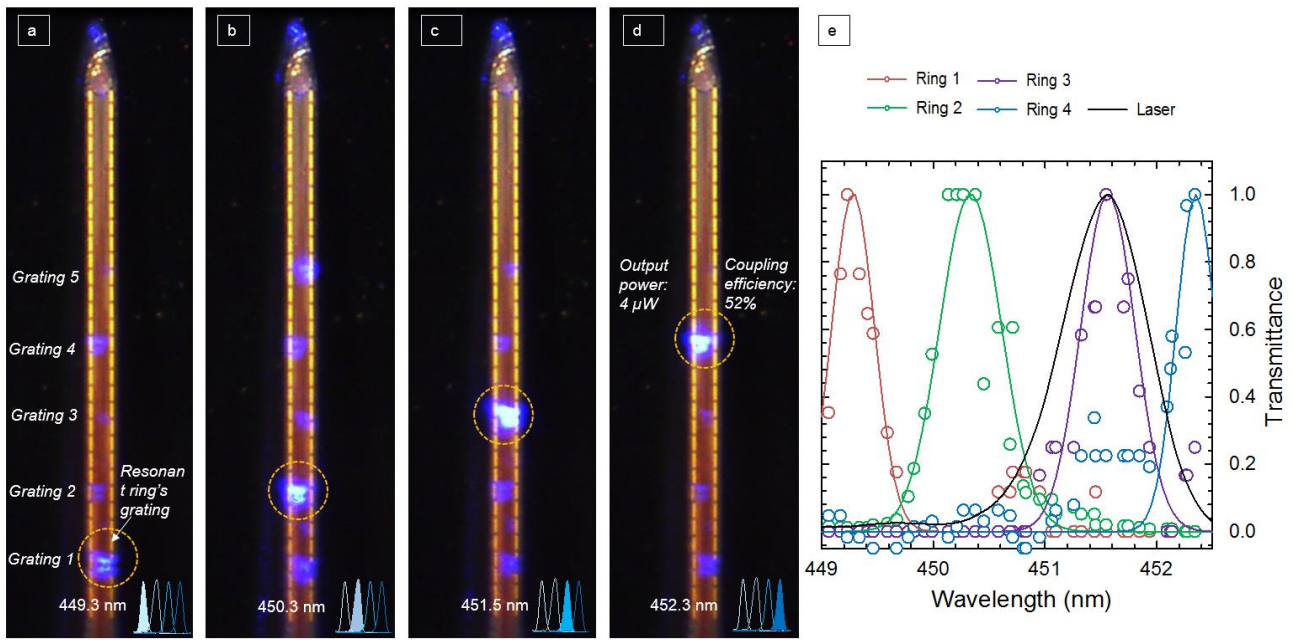

Figure 6. (a-d) optical microscope image of the tip of the assembled neural probe after we turn on the laser for the first four ring resonance wavelengths. In (d) we report the resonant grating's output

power and the fraction of total tip output power for a laser input of 10 mW (we obtain similar power results for (a-c)). All the gratings in the image are visible because we oversaturated it for graphical purposes. (e) Normalized experimental ring transmittance as a function of the laser input wavelength for different ring resonators and laser spectrum (in black). We fit the experimental output transmittance points with a gaussian curve.

**In vitro and *in vivo* experiments**

Next, we perform a preliminary *in* vivo test as probe's proof of concept, with the goal of verifying that we can simultaneously read neural activity across the entire tip and that we can optically stimulate neurons in selected vertical areas of interest. We select specific areas of interest by choosing specific output grating(s) and then matching the input laser wavelength to the corresponding ring's resonance. A mouse that had been previously implanted with a headplate and that showed good viral expression of the light-sensitive opsin ChR2 (more details in the Methods section) was selected for the *in vivo* characterization experiment. Cells expressing ChR2 simultaneously expressed TdTomato, a red fluorophore, allowing us to easily target the region with the brightest expression as our recording area. We recorded from the vibrissae Somatosensory cortex as an example of recording area (other areas can be analyzed as desired) and inserted the probe through a small (< 200 µm) craniotomy, lowered it to 1000 µm into the brain and let the electrode settle for five minutes before beginning our experiment.

Throughout the experiment, we record neural activity simultaneously across multiple electrodes- from which we identify and distinguish individual single units (presumptive neurons) and estimate their vertical positions along the probe, as shown in [28] (by data post-processing, see Methods section). Specifically, we divide the experiment into 14 blocks, each composed of 50 trials, which are three seconds long. We grouped the blocks into subgroups, each corresponding to an output grating. For each block in the subgroup, we chose a different input power, which changes the grating's output power proportionally. Within each subgroup, we observe that the laser input power modulates the neurons' firing rates (that is, the number of action potentials over time) increasing the rate with increasing laser power, as we show in Figure 7a. Firing rates increased when the light output of a specific grating is above a critical threshold. For a laser input value of 3 mW, corresponding to a grating output power of ~2 µW and a power density of around 40 mW/mm$^2$, we saw neural activity excitation.

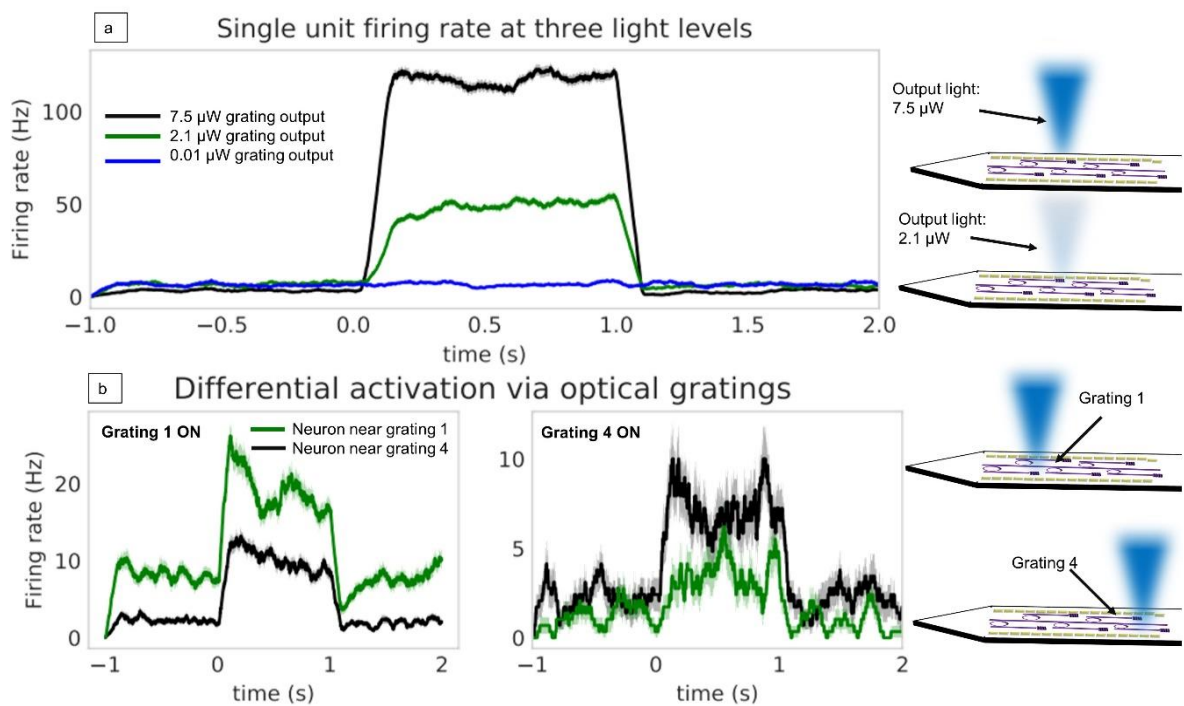

Figure 7. Preliminary result of the *in vivo* recording and simultaneous optogenetic excitation experiment performed in an optogenetically modified mouse. (a) Left panel: firing rate of a presumptive neuron (single unit) for different light output intensities (laser on between time 0 s to 1 s), firing rate increases proportionally for intensities above 2.1 µW (n=50 trials, mean +/- sem). Right panel: schematic illustration of the light output from the neural probe tip. (b) Excitation of neurons in different spatial locations based on the activated light output grating. Firing rates of two units when either grating 1 or grating 4 is on (left and center panels, respectively, n=50 trials, mean +/- sem) and schematic illustration of the probe's tip with different selected grating (right-hand side).

We found that the firing rates of neurons are modulated by light intensity. For a given optical grating we were able to modulate the firing rate of neighboring units depending on the light intensity being delivered. The optical sites evoked consistent neural activity as seen by the low standard error Figure 7a. More importantly, we found that we could modulate units' firing rate by selecting different gratings, as we show in Figure 7b. By selecting different optical gratings we caused units' firings rates to change dramatically, for a given light output, with some units increasing their rate and others decreasing their rate, presumably in proportion to their relative spatial position around the probe. Therefore, considering that the firing rates of putative units at unique depths did not change uniformly when we switched the light output location, we verified the ring's capability as a possible strategy to stimulate neural activity.

In conclusion, this experiment verified that our proof of concept optoelectric probe, integrating ring resonators and arrays of sensors, can both record and stimulate brain regions of interest by the tuning of the input wavelength and the selection of specific ring(s). These preliminary results show that our probe can be a viable technology for the scaling of passive optoelectrodes.

## Discussion

In this work, we realize and validate a neural probe designed with the simultaneous capability of *in vivo* electrophysiological recording and optogenetic manipulation of neural activity with high spatiotemporal resolution. While several devices can record and optically excite neural activity, here we make use for the first time of ring resonators to realize a compact probe combining a high density of sensors and stimulation sites, high scalability, and capability of addressing stimulation sites for on-demand manipulation of specific spatial regions without any significant heat generation.

Our design integrates the ring resonators into the nanophotonic circuit, where they act as optical switches that can be selected by tuning of an external laser's wavelength, thus providing an easy control on the capability of passively addressing single stimulation sites. Importantly, our nanophotonic design occupies a reduced lateral space on the tip thanks to both the small lateral footprint of the optical elements and the requirement of a single input waveguide for multiple outputs. In addition to the reduced lateral footprint, a careful design of the ring resonators allows for the increase of their finesse value, which ultimately corresponds to the number of independently addressable light outputs. This possibility is a paramount step towards a substantial increase in the density of light output sites, which is an unprecedented opportunity compared to state-of-the-art nanophotonic designs, with their need for one waveguide for each light output and therefore a stringent trade-off between the number of outputs and the overall tip dimensions. Moreover, we perform preliminary *in vivo* experiments in an optogenetically modified mouse, demonstrating the probe's working principle of locally stimulating spatial regions of interest in the brain, simultaneously to their recording.

In conclusion, this work shows the feasibility of microfabricating highly integrated, scalable and passively addressable neural probes that rely on a combination of arrays of sensors and nanophotonic circuits embedding ring resonators for scaling down neural optoelectrodes. As an example, our proof of concept device has a cross-sectional area coefficient (cross-section divided by the number of sensors and stimulation sites [23]) of 12, which is one order of magnitude smaller than the state of the art. Future directions include the optimization of our device towards its *in vivo* validation; access to a wider opsin selection by using different wavelengths in the visible range, as well as boosting the number of light output spots by increasing the ring finesse and an improved fabrication process throughput by nanoimprinting. Finally, we expect to optimize and use these devices to study the dynamics of the electrical signal propagation through the cortical layers in the somatosensory area by selectively silencing or exciting individual cortical layers.

## Materials and Methods

### Fabrication

We fabricate the neural probes using commercial wafers provided by Lionix that are made of silicon (525 µm thick), low pressure chemical vapor deposition $SiO_2$ (2.5 µm) and $Si_3N_4$ (160 nm); we perform all the fabrication processes at low temperature (<400 °C).

Alignment marks are initially patterned onto the substrate with electron beam lithography (using PMMA C4 resist), followed by electron beam evaporation of titanium and gold (10 nm and 100 nm respectively) and solvent liftoff (1 hour in Remover PG at 80°C).

The nanophotonic circuits and ring resonators are aligned to the marks and patterned with electron beam lithography (using ZEP 520A resist diluted at 50% and aquasave) and reactive ion etching (RIE, using $CHF_3/O_2$ chemistry with 48:2 gas ratio and forward power RF of 40 W). The process is repeatable across the wafer and was tested for multiple wafers, with minimal variations on the waveguide width that do not result in significant experimental performance change.

The nanophotonic circuits are optically insulated by depositing 2.7 µm of $SiO_2$ (with a plasma enhanced chemical vapor deposition tool, using 50 sccm of 1%$SiH_4$:Ar, 720 sccm of $N_2O$ and 160 sccm of $N_2$ at 150°C) and planarized by means of a flowable oxide (FOx 15 by Dow Corning, spun at 2000 rpm and baked on a hotplate at 350°C for 45 minutes).

The arrays of sensors are patterned with electron beam lithography (using 100% ZEP 520A), titanium/gold evaporation and liftoff (as for the patterning of the alignment marks). To passivate the wires, 60 nm thick $SiO_2$ is deposited with an atomic layer deposition tool (using a plasma and a temperature of 40°C), and then selectively removed from the electrodes with another electron beam exposure and RIE etching. We detail these processes in [28].

Trenches are then etched to define the probes' shape as well as the grooves for the optical fibers. To do this, we spin an optical lithography resist to mask the probe regions (AZ 40XT-11D, spun at 1750 rpm for 40 µm thickness). We use an *Oxford Plasma lab 100 Viper* tool to etch both the layers above the silicon (using 35 sccm $CF_4$, 15 sccm Ar and 10 sccm $O_2$, 150 W of RF, 400 W of very high-frequency power and 20°C of table temperature) and 15 µm of silicon (using 90 sccm $C_4F_8$ and 60 sccm $SF_6$ at 15 °C, 35 W RF, and 300 W VHF).

The silicon underneath the tip areas is then removed so as to make them thin (20 µm), while leaving the silicon underneath the probes' interfacing areas. To achieve this, we etch the silicon nitride on the wafer's backside with optical lithography (using resist MAP-1215) and RIE etching (same parameters as for the etching of the nanophotonic circuits). Both the map-1215 and the AZ-40XT-11D resists are removed with a 30-minute soak in AZ-400T.

Most of the silicon (480 µm) is removed from the previously nitride-etched areas on the wafer backside using potassium hydroxide (KOH) after a protective polymer coating on the circuits (Protek B3), as we describe in [28]. Besides, during the KOH etching we protect the wafer circuits by placing it in a wafer chuck (from AMMT). After the KOH etch, we remove the last few remaining µm of silicon with a dry etching in the VIPER tool (using $SF_6$ and $O_2$ chemistry).

**Ring resonator design**

The spacing between ring resonators depends on the illumination of interest and on the laser model, whose tunability range and FWHM limit the number of independently addressable ring resonators. The minimum spacing between rings can be as low as ~10 µm if designing the waveguide outputs perpendicular with respect to the bus (simulations show no difference between such configuration and the one shown in Fig. 3 and having waveguide outputs parallel with respect to the bus).

**Laser and optical fiber preparation**

*Laser:* we use a single-mode and fiber-coupled laser diode that is centered at the wavelength of 450 nm (model QFLD-450-10, from QPhotonics). We chose this small laser diode for a proof of concept testing of our optical system. Different laser models, having a wider tunability range, faster switching capabilities and longer temporal stability can be chosen for future in vivo experiments. The diode has a maximum output power of around 10 mW. We tune the laser's wavelength by changing the temperature, through the laser controller, in the range 10°C to 60°C and monitor the corresponding wavelength with a spectrometer, as we describe in the next paragraph. We perform an initial calibration as we show in Fig. 8

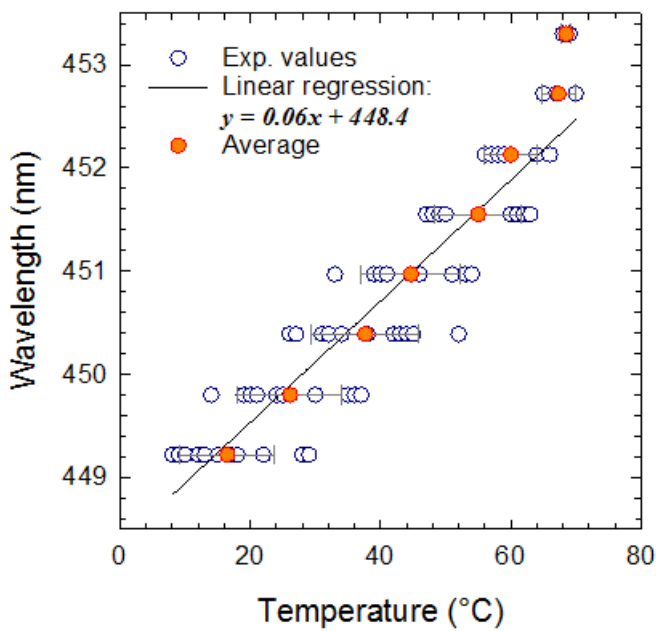

Fig. 8: Laser diode temperature-wavelength calibration curve

*Probe coupled fiber:* we use single-mode optical fibers (model SM 400 from Thorlabs). We thin one side of the fiber such that it fits inside the 20 µm deep groove using either mechanical polishing or hydrofluoric acid etching. The first procedure is described in [36], [37] and allows for the mechanical removal of ~40 µm of the fiber's cladding on one side. The second one is performed by immersion of the fiber in >40% HF for 45 minutes, followed by careful cleaning. Such fiber is then spliced to a patch cord so that it can be readily connected to the laser's one.

*Ring resonator experimental measurement:* we measure the ring resonators' spectral responses by extracting their corresponding grating output intensities while changing the laser's wavelength through its temperature control.

Since the laser's wavelength is a nonlinear function of its temperature and its output power varies for different temperatures, we implement a feedback loop with a Matlab script that adjusts both the wavelength to match the desired one (by tuning the temperature) and the laser output power to a constant value across every wavelength (by changing its input current).

**Device optical losses**

The laser diode we use for the experiment has a maximum output power of around 10 mW. We estimate the total system losses to be around 30 dB. Of these, 1 to 2 dB are due to the laser fiber-probe fiber FC/PC connection. 10-15 dB losses are due to the edge coupling (10 dB) and fiber gluing (around 5 dB), 6-8 dB waveguide transmission losses (due to light scattering on the waveguide's sidewalls). 3 dB losses are due to the grating outcoupling losses [30]. Besides, rings couple only 45-60% of the input light when their resonance frequency is matched (this introduces 3 dB additional losses). The total output power is, therefore, 5 to10 uW, which corresponds to a power density of ~ 100 mW/mm2 if we consider a light output spot dimension on the order of the grating dimension (5 um X 10 um). The power is calculated from the microscope images by calibrating the CCD camera; we verify these calculations by comparing the probe output light measured with a powermeter.

The system optical losses could be drastically reduced by further optimization: for example propagation losses for nitride waveguides were reported to be below 1 dB/cm at 532 nm wavelength [38], and coupling and gluing losses could be reduced by using i.e. grating coupling methods (~4.8 dB from our FDTD simulations) and by a more careful alignment (<3 dB), thus lowering the total losses from 30 dB to below 15 dB. Furthermore, most of the losses are at the fiber-waveguide interface in the probe interface area, which is far from the tip.

**Gratings**

We simulate gratings with FDTD and use the following parameters: grating length and width 15 X 4 µm, pitch 315 nm (for emission at 16° for 450 nm wavelength), and duty cycle 0.5. Different grating designs could be chosen depending on the application of interest (i.e. focusing as we described in [28] or long-distance collimation [12]) but more detailed tests will be performed in future studies.

*In vivo* **experiment**

All experiments involving mice were performed in the Adesnik Lab, UC Berkeley, in accordance with the guidelines and regulations of the Animal Care and Use Committee (Protocol # AUP-2014-10-6832-1). The mice used in these experiments were wild type (CD-1, Charles River Laboratories) and underwent two surgical procedures in preparation for the *in vivo* optoelectrode tests. Briefly, in the first procedure mice were anesthetized using 2% isoflurane and head-fixed to a stereotax using proper aseptic technique. The scalp and the underlying fascia were removed to expose the dorsal part of the skull. Two small craniotomies were made using a dental drill, one over vibrissae primary somatosensory cortex (vS1) and the other over vibrissae motor cortex (vM1). A microinjector was used to inject 400uL of an adeno-associated virus (AAV), carrying a genetic payload that causes infected neurons to produce Cre-recombinase (Cre), into vS1. An additional injection of 400uL of a second AAV carrying a payload that causes neurons to express the excitatory ion channel Channelrhodopsin (ChR2) in neurons that also contain Cre was injected into vM1. These two brain regions share many reciprocal connections giving us a large target area to test the probes. Once the injections were complete the entire skull was covered in Vetbond (3M) to seal the wound margins and protect the skull. A custom aluminum headplate was attached to the skull using dental acrylic (Metabond). The mice were then taken off isoflurane and allowed to recover. These mice were given a week to acclimatize to the new headplate as well as being head-fixed to the rotary treadmill where the experiments would take place.

The second procedure was conducted on the day of the experiment. Here, previously injected and heaplated mice that showed good expression of the excitatory opsin were anesthetized using 2% isoflurane. A small dental drill was used to thin the skull over the region of the brightest expression. After thinning a 27g needle was used to lift a small flap of the skull to expose the brain. Mice were recovered from anesthesia and placed on the rotary treadmill in the electrophysiology rig. Here the electrode was fastened to a micromanipulator (Sensapex) and lowered roughly 1000um into the brain, ensuring all electrodes were in the cortex. All neural recordings were conducted at a sampling rate of 30kHz and recorded with SpikeGadgets hardware and software. During the recording, neural activity was clearly present and modulated with the activation of the light pads on the probe.

Postprocessing of the data was conducted with custom MATLAB (Mathworks) and Python software. Semi-automated spike sorting was conducted using the freely available Klusta, which uses a custom sorting algorithm that takes probe geometry into account to identify spike times associated with specific units.

## Acknowledgments
The authors would like to thank Andrea Lamberti and Candido Pirri (Department of Applied Science and Technology, Politecnico di Torino) for supervision. Fabrizio Riminucci for advice on fiber-waveguide coupling FDTD simulations. Phathakone Sanethavong for wire bonding work. Kristofer Bouchard for helpful advice on the in vitro tests. Stefano Dallorto for helpful advice on etching processes. Arian Gashi for proofreading the manuscript.

Work at the Molecular Foundry was supported by the Office of Science, Office of Basic Energy Sciences, of the U.S. Department of Energy under Contract No. DE-AC02-05CH11231.

## Conflict of interests

The Authors declare no Conflict of Interest. The work has not been published or submitted for publication elsewhere and no materials are reproduced from another source.

## Author contributions

V. Lanzio was the main researcher working on this project and designed, simulated, fabricated, optimized, assembled and experimentally characterized in vitro the neural probes. G. Telian performed the *in vivo* experiment and corresponding data extraction, supervised by H. Adesnik. A. Koshelev conceptualized the nanophotonic circuit and helped in initial testing, followed by simulations and testing performed by V. Lanzio and P. Micheletti. G.Presti, E. D'Arpa and P. De Martino helped in the optimization of all the fabrication processes, supervised by V. Lanzio. S. Dhuey performed the electron beam lithography steps. The manuscript writing, data interpretation and discussion was done by V. Lanzio with the help of M. Lorenzon. The project was conceptualized by S. Cabrini, H. Adesnik and P. Denes. S. Sassolini and M. West supervised the initial part of the project, while S. Cabrini supervised the entire project.